# Impact of impurities on zonal flow driven by trapped electron mode turbulence

Weixin Guo, Lu Wang[1] and Ge Zhuang

State Key Laboratory of Advanced Electromagnetic Engineering and Technology, School of Electrical and Electronic Engineering, Huazhong University of Science and Technology, Wuhan, Hubei 430074, China

E-mail: luwang@hust.edu.cn

## Abstract

The impact of impurities on the generation of zonal flow (ZF) driven by collisonless trapped electron mode (CTEM) turbulence in deuterium (D)-tritium (T) plasmas is investigated. The expression for ZF growth rate with impurities is derived by balancing the ZF potential shielded by polarization effects and the ZF modulated radial turbulent current. Then, it is shown that the maximum normalized ZF growth rate is reduced by the presence of the fully ionized non-trace light impurities with relatively flat density profile, and slightly reduced by highly ionized trace tungsten (W). While, the maximum normalized ZF growth rate can be also enhanced by fully ionized non-trace light impurities with relatively steep density profile. In particular, the effects of high temperature helium from D-T reaction on ZF depend on the temperature ratio between electron and high temperature helium. The possible relevance of our findings to recent experimental results and future burning plasmas is also discussed.

## I. Introduction

Zonal flow (ZF) is widely believed to play a significant role in suppressing turbulence and reducing anomalous transport level [1], and thus is important for L-H transition in magnetic fusion plasmas. ZF can be excited by micro-turbulence, such as ion temperature gradient (ITG), trapped electron mode (TEM) and electron temperature gradient (ETG) through the nonlinear interaction. In particular, TEM turbulence, which is a promising candidate responsible for anomalous electron heat transport has been reported in Alcator C-Mod [2], ASDEX upgrade [3], and DIII-D [4] when powerful electron heating is applied. The TEM driven turbulence and the associated transport may be also important in future burning plasmas such as International

---
[1] Author to whom any correspondence should be addressed.

Thermonuclear Experimental Reactor (ITER) and Demonstration Power Plants (DEMO) where the energetic alpha particles from deuterium (D)-tritium (T) reaction dominantly heat electrons [5, 6]. Moreover, the anomalous electron heat transport is still observed in experiments although ITG is suppressed and the ion heat transport is effectively reduced. Therefore, it is worth studying ZF driven by TEM turbulence especially when ITG mode is stabilized. Actually, the nonlinear gyrokinetic simulations have studied the ZF in the collisonless trapped electron mode (CTEM) turbulence [7-9], and the analytical calculation of ZF growth rate from TEM turbulence was also reported [10] basing on the modulational theory [11].

In addition, *impurities* are inevitable components in fusion plasmas because of the interaction between main plasmas and wall material. Helium ash from D-T reaction is an important impurity source as well. Impurities with finite concentration (the ratio of impurity density to electron density) are non-negligible in the quasi-neutrality condition and may influence the turbulence and turbulent transport, thus, the usual trace approximation [12-14] is failed. This kind of impurity with finite concentration is called as the non-trace impurity [15, 16]. Actually, there has been abundant theoretical and numerical works studying the impact of impurities on ITG [17-20], TEM [21-23] and electromagnetic (EM) [24, 25] turbulence. The experimental observations in DIII-D found the enhancement of **E** x **B**$_t$ shearing rate and the confinement improvement by injecting Neon (Ne) [26]. The influences of highly charged impurities on collisional damping of large scale ZF [27] were analytical derived in [28]. Moreover, the impact of various impurities on arbitrary scale collisionless residual ZF through the modification of polarization shielding is systematically investigated in our previous work [16]. But, studying the *effects of impurity* on ZF generation from conjunctively considering the effects of impurities on both polarization shielding and on turbulence, to our knowledge, has not been investigated. Furthermore, the simulation works in JET found the highest fusion yields in approximately 1:1 D-T mixture [29], and ZF is predicted to play an possibly important role in this process. Therefore, comprehensive studying impurity effects on the generation of ZF driven by CTEM turbulence in D-T plasmas is of great significance.

In this work, we explore how impurities with the tolerance concentration in ITER and JET affect ZF driven by CTEM turbulence in D-T plasmas. The expression for ZF growth rate including the impurity polarization shielding and the impurity effects on CTEM instability is derived. Especially, ZF modulated radial current driven by CTEM turbulence is directly calculated, which is consistent with previous results in Refs. [10, 11]. The principal results of the present paper are as follows.

(1) Impurities with inwardly (outwardly) peaked density profile stabilize (destabilize) the CTEM instability when electron temperature is comparable to ion/impurity temperature, and these effects are strengthened by steepening the impurity density profile.

(2) The normalized ZF growth rate is inversely proportional to the total polarization shielding, which includes the components of both ion and impurity. Meanwhile, the maximum normalized ZF growth rate is also found to be proportional to $\left(\frac{\hat{\gamma}_k}{|\hat{v}_{gr}|}\right)^3$ with $\hat{\gamma}_k$ and $\hat{v}_{gr}$ being the normalized CTEM growth rate and the normalized radial group velocity, respectively. Therefore, impurity effects on both polarization shielding and on $\frac{\hat{\gamma}_k}{|\hat{v}_{gr}|}$ should be combined to evaluate the impurity effects on ZF.

(3) The fully ionized non-trace impurities with relatively flat density profile, either inwardly peaked or outwardly peaked, reduce the maximum normalized ZF growth rate. This is mainly because the enhancement of ZF shielding effects by the presence of impurities is stronger than impurity effects on $\frac{\hat{\gamma}_k}{|\hat{v}_{gr}|}$. While, when the impurity density profile becomes steeper, the fully ionized non-trace impurities can also enhance the maximum normalized ZF growth rate. This is because the enhancement of $\frac{\hat{\gamma}_k}{|\hat{v}_{gr}|}$ overcomes the enhancement of ZF polarization shielding. In addition, the maximum normalized ZF growth rate is slightly reduced by highly ionized trace tungsten (W) due to slightly enhancement of the total polarization shielding.

(4) Especially, the effects of high temperature helium ($He^{2+}$) impurity from D-T reaction on the normalized ZF growth rate are dependent on the temperature ratio between electron and $He^{2+}$.

The remainder of this paper is organized as follows. In Sec. II, we derive the expression for ZF growth rate driven by CTEM turbulence in the presence of impurities. In Sec. III, the effects of impurity on both polarization shielding and on CTEM instability are analyzed, respectively. Then, the impact of impurities on ZF growth rate in D-T plasmas is systematically investigated in Sec. IV. Finally, the conclusions and discussions are in Sec. V.

## II. Analytical expression for ZF growth rate in the presence of impurities

In this section, we present the derivation of ZF growth rate driven by CTEM turbulence. Both the effects of impurities on both polarization shielding and CTEM instability are included. According to [11], the total radial current including both the polarization current and the ZF modulated turbulent current maintains divergence-free condition

$$\nabla \cdot \left( \vec{J}_p + \langle \vec{J}_r \rangle_{ZF} \right) = 0 , \quad (1)$$

where $\vec{J}_p$ and $\langle \vec{J}_r \rangle_{ZF}$ are the polarization current and ZF modulated turbulent current, respectively, with $\langle ... \rangle$ representing the fast spatial and temporal scale averaged operation over the phase of CTEM. Using the continuity equation of the polarization density [30], we can obtain the relationship between ZF potential shielded by the polarization effects and the ZF modulated turbulent current

$$-en_{0e}\left( g_i \chi_{i,z} + g_z \chi_{z,z} \right) \frac{\partial \Phi_{ZF}}{\partial t} = \frac{\partial}{\partial r} \langle J_r \rangle_{ZF}. \quad (2)$$

Here, $e$ is the elementary charge, $g_i = (1 - Zf_c)\tau_i$ and $g_z = Z^2 f_c \tau_z$ are weighting factors corresponding to ion and impurity, respectively, $Z$ is the charge number of impurity, $f_c = n_{0z}/n_{0e}$ with $n_{0z}$ and $n_{0e}$ being the impurity and electron equilibrium densities, $\tau_i = T_e/T_i$ and $\tau_z = T_e/T_z$ are temperature ratios with $T_\alpha$ being the temperature and the index $\alpha = e, i, z$ corresponding to electron, ion and impurity, respectively, $\chi_{i,z}$ and $\chi_{z,z}$ represent the generalized ion and impurity polarization shieldings of the arbitrary scale ZF potential, respectively, $\Phi_{ZF} = e\delta\phi_{ZF}/T_e$ with $\delta\phi_{ZF}$ being the ZF potential.

In the following, we will directly calculate the slow spatiotemporal variational ZF modulated turbulent current $\langle J_r \rangle_{ZF}$. Assuming the wave vector and frequency of ZF are $\vec{q} = (q_r, 0, 0)$ and $\Omega$, respectively, $\langle J_r \rangle_{ZF}$ can be calculated as

$$\langle J_r \rangle_{ZF} = en_{0e}\left(\langle(1-Zf_c)\hat{n}_i\tilde{v}_r\rangle + \langle Zf_c\hat{n}_z\tilde{v}_r\rangle - \langle \hat{n}_e\tilde{v}_r\rangle\right)$$

$$= en_{0e}\left\langle \begin{array}{l} \sum_{\vec{k}+\vec{q}}\chi(\vec{k}+\vec{q},\omega+\Omega)\Phi_{\vec{k}+\vec{q},\omega+\Omega}\exp\left[i(\vec{k}+\vec{q})\cdot\vec{x}-i(\omega+\Omega)t\right] \\ \times \sum_{\vec{k}'}-ik'_\theta c_s\rho_s \Phi_{\vec{k}',\omega'}\exp(i\vec{k}'\cdot\vec{x}-i\omega' t) \end{array}\right\rangle$$

$$= en_{0e}\sum_{\vec{k},\omega}\sum_{\vec{q},\Omega}\chi(\vec{k}+\vec{q},\omega+\Omega)ik_\theta c_s\rho_s |\Phi_k^D|^2 \exp(i\vec{q}\cdot\vec{x}-i\Omega t) \quad . \tag{3}$$

Here, $\vec{k}$ and $\omega$ are the wave vector and frequency of CTEM and we assume that $q_r \ll k_r$, $\Omega \ll \omega$ with $k_r$ being the radial wavenumber of CTEM. $|\Phi_k^D|^2 = \Phi_{\vec{k}+\vec{q},\omega+\Omega}\Phi^*_{\vec{k},\omega}$ ($\Phi_k = e\delta\phi_k/T_e$ with $\delta\phi_k$ being the electric potential fluctuation of CTEM) is the weakly modulated inhomogeneous spectral of drift wave (DW) fluctuations [31]. The other symbols are: $\hat{n}_\alpha = \delta n_\alpha/n_{0\alpha}$ ($\alpha = e, i, z$) is the normalized density with $\delta n_\alpha$ and $n_{0\alpha}$ being the perturbed and the equilibrium densities, $\tilde{v}_r = \sum_k -ik_\theta c_s \rho_s \Phi_k$ is the fluctuating radial **E** x **B** velocity with $k_\theta$ being the poloidal wavenumber of CTEM, $c_s = \sqrt{T_e/m_i}$ with $m_i$ being the ion mass, $\rho_s = c_s/\Omega_i$ with $\Omega_i = eB_t/(cm_i)$ being the ion cyclotron frequency, $B_t$ being the total magnetic field, $c$ being the light velocity. The total susceptibility can be written as,

$$\chi(\vec{k}+\vec{q},\omega+\Omega) = g_i\chi_i + g_z\chi_z - \chi_e \tag{4}$$

with $\chi_\alpha$ being the susceptibility of species, i.e., $\hat{n}_\alpha = \chi_\alpha \dfrac{Z_\alpha e\delta\phi_{\vec{k}+\vec{q}}}{T_\alpha}$ with $Z_\alpha$ being the charge number of species ($Z_\alpha = Z$ for impurities; $Z_\alpha = 1$ for both ions and electrons) and $T_\alpha$ being the temperature. Expanding $\chi(\vec{k}+\vec{q},\omega+\Omega)$ as

$$\chi(\vec{k}+\vec{q},\omega+\Omega) = \chi(\vec{k},\omega) + \Omega\frac{\partial\chi}{\partial\omega} + q_r\frac{\partial\chi}{\partial k_r} . \tag{5}$$

According to the local quasi-neutrality condition, the lowest order satisfies

$$\chi(\vec{k}, \omega) = 0. \tag{6}$$

Then, taking Eqs. (5), (6) into Eq. (3) gives

$$\langle J_r \rangle_{ZF} = en_{0e} \sum_{\vec{k},\omega} \sum_{q,\Omega} (\Omega \frac{\partial \chi}{\partial \omega} + q_r \frac{\partial \chi}{\partial k_r}) ik_\theta c_s \rho_s |\Phi_k^D|^2 \exp(i\vec{q}\cdot\vec{x} - i\Omega t). \tag{7}$$

Using the perturbed "pseudo-action" density defined as $\tilde{N}_k = -\frac{n_{0e}T_e}{2}\frac{\partial \chi}{\partial \omega}|\Phi_k^D|^2$ in [31], Eq. (7) can be reduced as

$$\langle J_r \rangle_{ZF} = 2\sum_{\vec{k},\omega}\sum_{q,\Omega} -i\frac{ck_\theta}{B_t}\left[(\Omega + v_{gr}q_r)\tilde{N}_k\right]\exp(i\vec{q}\cdot\vec{x} - i\Omega t), \tag{8}$$

where $v_{gr} = \frac{\partial \omega}{\partial k_r}$ is the radial group velocity of DW. Note that, it can be easily demonstrated that the directly calculated $\langle J_r \rangle_{ZF}$, i.e., Eq. (8) is equal to the modulated turbulent driven radial current $\frac{\delta J_r}{\delta \Phi_{ZF}}\Phi_{ZF}$ in [10] under the condition that ZF growth rate is much smaller than the linear growth rate of CTEM.

The evolution of the total DW "pseudo-action" density $N_k$ follows [31]

$$\frac{\partial N_k}{\partial t} + (\vec{v}_g + \vec{v}_{ZF})\cdot\frac{\partial N_k}{\partial \vec{r}} - \frac{\partial}{\partial \vec{r}}(\vec{k}\cdot\vec{v}_{ZF})\cdot\frac{\partial N_k}{\partial \vec{k}} = \Gamma_k N_k - \frac{\Delta\omega_k}{N_0}N_k^2. \tag{9}$$

Here, $N_k = \langle N_k \rangle + \tilde{N}_k$ with $\langle N_k \rangle$ being the magnetic surface averaged "pseudo-action" density, $\vec{v}_{ZF} = \frac{cT_e}{B_t}\vec{b}\times\nabla\Phi_{ZF}$ represents the ZF velocity with $\Phi_{ZF} = \sum_{q,\Omega}\Phi_{ZF,q_r}e^{i\vec{q}\cdot\vec{x}-i\Omega t}$ and $\vec{b}$ being the unity vector along the magnetic field, $\Gamma_k$ refers to the linear increment of DW, and the equilibrium spectrum $N_0 \approx \langle N_k \rangle$, $\Delta\omega_k$ represents the nonlinear damping of DW. Linearizing Eq. (9) and assuming $\Delta\omega_k \approx \Gamma_k$, we can obtain the perturbed "pseudo-action" density $\tilde{N}_k$

$$\tilde{N}_k = -iq_r^2 \frac{k_\theta c_s \rho_s}{\Omega - q_r v_{gr} + i\Gamma_k}\frac{\partial \langle N_k \rangle}{\partial k_r}\Phi_{ZF,q_r}. \tag{10}$$

Now, we substitute Eqs. (8) and (10) into Eq. (2) and obtain the frequency of ZF

$$\Omega = \frac{2q_r}{en_{0e}\left(g_i\chi_{i,z} + g_z\chi_{z,z}\right)} \sum_{\bar{k},\omega} -\frac{ck_\theta}{B_t} q_r^2 k_\theta c_s \rho_s \frac{\Omega + v_{gr}q_r}{\Omega - q_r v_{gr} + i\Gamma_k} \frac{\partial \langle N_k \rangle}{\partial k_r}. \quad (11)$$

For zero real frequency ZF, i.e., $\Omega = i\gamma_{ZF}$ with $\gamma_{ZF}$ being the growth rate of ZF. Finally, we have

$$\gamma_{ZF} \approx \frac{2q_r^4 c_s^2 \rho_s^2}{n_{0e}T_e\left(g_i\chi_{i,z} + g_z\chi_{z,z}\right)} \sum_{\bar{k},\omega} k_\theta^2 \frac{\Gamma_k v_{gr}}{\left(q_r v_{gr}\right)^2 + \Gamma_k^2} \frac{\partial \langle N_k \rangle}{\partial k_r}. \quad (12)$$

The linear increment term $\Gamma_k$ can be expressed as $\Gamma_k = 2\gamma_k$ according to [31, 32], where $\gamma_k$ is the linear growth rate of CTEM. In the process of deriving Eq. (12), the approximation of $\gamma_{ZF} \ll \gamma_k$ was used, which is usually adopted in the problem of disparate scale interaction. The radial group velocity $v_{gr}$ is negative for the usual electron DW ($k_r > 0$ is taken), and $\frac{\partial \langle N_k \rangle}{\partial k_r} < 0$, i.e., monotonically decaying spectrum has been observed in experiments [33, 34]. Therefore, we can expect positive ZF growth for $v_{gr}\frac{\partial \langle N_k \rangle}{\partial k_r} > 0$. In addition, dividing this term into negative radiation pressure (driving force) and positive radiation pressure (retarding force) and the competition between them to drive seed ZF were discussed in [35].

In the following, we use the ballooning representation in the toroidal configuration and consider the resonant trapped electrons and the adiabatic passing electrons. Thus, the susceptibility of electrons can be calculated [36]

$$\chi_{e,k} = 1 - \sqrt{2\varepsilon_0}\left(1 - \frac{\omega_{*e}}{\omega}\right) - \frac{3}{2}\sqrt{2\varepsilon_0}\frac{\overline{\omega}_{de}G_{av}}{\omega}\left[1 - \frac{\omega_{*e}}{\omega}(1+\eta_e)\right]$$

$$+ i2\sqrt{2\pi\varepsilon_0}\left(\frac{\omega}{\overline{\omega}_{de}G_{av}}\right)^{3/2} \exp\left[-\frac{\omega}{\overline{\omega}_{de}G_{av}}\right]\left\{1 - \frac{\omega_{*e}}{\omega}\left[1+\eta_e\left(\frac{\omega}{\overline{\omega}_{de}G_{av}} - \frac{3}{2}\right)\right]\right\}. \quad (13)$$

Here, $\varepsilon_0$ is the inverse aspect ratio, $\omega_{*e} = \frac{k_\theta}{L_{ne}}\frac{cT_e}{eB_t}$ is the electron diamagnetic frequency with $L_{ne} = -\frac{n_{0e}}{\nabla n_{0e}}$ being the electron density gradient scale length, $\overline{\omega}_{de}$ is the bounce averaged trapped electron magnetic precession frequency, and the averaged value $G_{av}$ over the azimuthal angle is expressed as $G_{av} = 0.64\hat{s} + 0.57$ [37] with $\hat{s} = \frac{r}{q}\frac{dq}{dr}$ being the magnetic shear, $r$ being the radial direction and $q$ being the safety factor, $\eta_e = L_{ne}/L_{Te}$ with $L_{Te} = -\frac{T_e}{\nabla T_e}$ being the electron temperature gradient scale length. For ions and impurities, we neglect the ballooning and resonant effects by assuming sufficiently radially localized modes near the rational surfaces as in [36]. Then, the susceptibility of ions or impurities is obtained by taking the lowest order hydrodynamic approximation,

$$\chi_{i(z),k} = -\left\{1 - \Gamma_{0i(z)} + \frac{\omega_{*i(z)}}{\omega}\left[\Gamma_{0i(z)} + \eta_{i(z)}b_{i(z)}\left(\Gamma_{1i(z)} - \Gamma_{0i(z)}\right)\right]\right\}. \quad (14)$$

Here, $\Gamma_{ni(z)} = I_n(b_{i(z)})\exp(-b_{i(z)})$, $I_n$ being the *nth* order modified Bessel function, $b_{i(z)} = k_\perp^2 \rho_{i(z)}^2$, and $\rho_{i(z)}$ being the gyroradius of ions or impurities, $\omega_{*i} = -\frac{k_\theta}{L_{ni}}\frac{cT_i}{eB_t}$ and $\omega_{*z} = -\frac{k_\theta}{L_{nz}}\frac{cT_z}{ZeB_t}$ being the ion and impurity diamagnetic frequencies, respectively, $L_{ni(z)} = -\frac{n_{0i(z)}}{\nabla n_{0i(z)}}$ being the density gradient scale length of ions (impurities), $\eta_{i(z)} = L_{ni(z)}/L_{Ti(z)}$ with $L_{Ti(z)} = -\frac{T_{i(z)}}{\nabla T_{i(z)}}$ being the temperature gradient scale length of ions (impurities). Generally speaking, $L_{n\alpha}$ as the density scale length of electrons, ions or impurities is positive for inwardly peaked profile. But, *specially,* $L_{nz}$ can be negative when impurities are outwardly peaked (i.e., $\nabla n_{0z} > 0$). Taking Eqs. (13) and (14) into the quasi-neutrality equation, we obtain the real frequency and the linear growth rate of CTEM instability,

$$\hat{\omega}_r = \frac{1}{\Delta}, \quad (15)$$

$$\hat{\gamma}_k = \frac{1}{\Delta^2 B}2\sqrt{2\pi\varepsilon_0}\left(\frac{R\hat{\omega}_r}{L_{ne}G_{av}}\right)^{3/2}\exp\left[-\frac{R\hat{\omega}_r}{L_{ne}G_{av}}\right]\left\{\frac{1}{\hat{\omega}_r}\left[1 + \eta_e\left(\frac{R\hat{\omega}_r}{L_{ne}G_{av}} - \frac{3}{2}\right)\right] - 1\right\}. \quad (16)$$

Here, $\hat{\omega}_r = \omega_r/\omega_{*e}$ and $\hat{\gamma}_k = \gamma_k/\omega_{*e}$ are the normalized frequency and linear growth rate of CTEM instability, $\Delta = -\frac{A}{B}$ with $A = -\left[1 - \sqrt{2\varepsilon_0} + (1-\Gamma_{0i})g_i + (1-\Gamma_{0z})g_z\right]$, $B = (1-Zf_c)L_{ei}\left[\Gamma_{0i} + \eta_i b_i(\Gamma_{1i} - \Gamma_{0i})\right] + Zf_c L_{ez}\left[\Gamma_{0z} + \eta_z b_z(\Gamma_{1z} - \Gamma_{0z})\right] - \sqrt{2\varepsilon_0}\left(1 - \frac{3}{2}G_{av}\frac{L_{ne}}{R}\right)$, $L_{ei} = L_{ne}/L_{ni}$, $L_{ez} = L_{ne}/L_{nz}$, $R$ being the major radius. The relationship $(1-Zf_c)L_{ei} + Zf_c L_{ez} = 1$ is indicted from the equilibrium quasi-neutrality condition. If we further assume $R/L_{Tz} = R/L_{Ti}$, we can also have $\eta_z = \eta_i \frac{\frac{1}{L_{ez}} - Zf_c}{1 - Zf_c}$. It should be noted that the coefficients of $A$ and $B$ can be reduced to the same expression in [10] in the absence of impurities.

Moreover, the ZF growth rate in Eq. (12) can be explicitly expressed if the spectrum of DW is given. For simplicity, we employ a monochromatic wave packet, i.e., the equilibrium "pseudo-action" density is represented as $\langle N_k \rangle = N_0 \delta(\vec{k} - \vec{k}_0)$ with the two-dimensional wave number $\vec{k}_0 = (k_{r0}, k_{\theta 0})$ [38], where the condition for ZF growth, $N_0 \frac{\partial v_{gr}}{\partial k_{r0}} < 0$, was pointed out, and it is usually satisfied for typical electron DW. This is similar to the so-called Lighthill criterion [39] for modulational instability [40]. We approximate the summation in $\vec{k}$ in Eq. (12) by an integration and then integrate by parts, and $\partial \left[\frac{\Gamma_{k_0} v_{gr}}{(q_r v_{gr})^2 + \Gamma_{k_0}^2}\right] / \partial k_{r0} < 0$ is required for positive ZF growth. Then, the ZF growth rate which is normalized to $\frac{v_{thi}}{R}\left(\frac{R^2}{\rho_i^2}\right)\left(\frac{e\delta\phi_{k_0}}{T_i}\right)^2$ can be unambiguously written as

$$\hat{\gamma}_{ZF} = \frac{2q_r^4 \rho_i^4}{\tau_i(g_i \chi_{i,z} + g_z \chi_{z,z})} \frac{\hat{\gamma}_k}{|\hat{v}_{gr}|} \frac{L_{ne}}{R} B\Delta^2 \left[1 + 2k_{r0}^2 \rho_i^2 \left(\frac{\Delta''}{\Delta'} - 2\frac{\Delta'}{\Delta}\right)\right] \frac{4\frac{\hat{\gamma}_k^2}{\hat{v}_{gr}^2} - q_r^2 \rho_i^2}{\left(4\frac{\hat{\gamma}_k^2}{\hat{v}_{gr}^2} + q_r^2 \rho_i^2\right)^2}. \quad (17)$$

Here, $\hat{v}_{gr} = -2k_{r0}\rho_i \Delta'/\Delta^2$ with $\Delta'$ and $\Delta''$ representing the first and second order of differential operation with respect to $b_{i(z)}$. Especially, we choose $k_{r0} > 0$ and use $k_{r0} = k_{\theta 0}$ in the derivation process, then $\hat{v}_{gr}$ is negative due to $\Delta' > 0$ ($\Delta' = \dfrac{A\dfrac{\partial B}{\partial b_i} - B\dfrac{\partial A}{\partial b_i}}{B^2} + \dfrac{A\dfrac{\partial B}{\partial b_z} - B\dfrac{\partial A}{\partial b_z}}{B^2}\dfrac{b_z}{b_i}$, it is very easy to find out that $A<0$, $B>0$, $\dfrac{\partial B}{\partial b_{i(z)}} < 0$ and $\dfrac{\partial A}{\partial b_{i(z)}} < 0$). This is why we use the absolute value $|\hat{v}_{gr}|$ in the expression of $\hat{\gamma}_{ZF}$. Therefore, the region of ZF radial scale corresponding to ZF growth can be deduced from Eq. (17) as $0 < q_r \rho_i < 2\dfrac{\hat{\gamma}_k}{|\hat{v}_{gr}|}$.

Moreover, Eq. (17) also shows that the generation of ZF can be affected by impurity through the impurity effects on both CTEM instability and the total polarization shielding, i.e., $g_i \chi_{i,z} + g_z \chi_{z,z}$.

In Eq. (17), the normalized ZF growth rate depends on the ZF scale and can be reflected from $q_r^4 \rho_i^4 \dfrac{4\dfrac{\hat{\gamma}_k^2}{\hat{v}_{gr}^2} - q_r^2 \rho_i^2}{\left(4\dfrac{\hat{\gamma}_k^2}{\hat{v}_{gr}^2} + q_r^2 \rho_i^2\right)^2}$. We can easily find that the variation of the normalized ZF growth rate with ZF scale is non-monotonic, i.e., $\hat{\gamma}_{ZF}$ increases with the increase of $q_r^2 \rho_i^2$ and then decreases with it. Through $\dfrac{\partial \hat{\gamma}_{ZF}}{\partial (q_r \rho_i)} = 0$, we can get the maximum normalized ZF growth rate, $\hat{\gamma}_{ZF,\max}$ at $q_r^2 \rho_i^2 = \left(2\sqrt{17} - 6\right)\left(\dfrac{\hat{\gamma}_k}{\hat{v}_{gr}}\right)^2$. Then, the *main* scaling of the maximum normalized ZF growth rate can be written as

$$\hat{\gamma}_{ZF,\max} \sim \dfrac{\left(\dfrac{\hat{\gamma}_k}{|\hat{v}_{gr}|}\right)^3}{g_i \chi_{i,z} + g_z \chi_{z,z}}. \tag{18}$$

Therefore, the overall effects of impurities on ZF generation is determined by the combination of the impurity effects on polarization shielding and on $\frac{\hat{\gamma}_k}{|\hat{v}_{gr}|}$, which will be presented in section III and IV.

## III. Impurity effects on polarization shielding and on CTEM instability.

Before showing the effects of impurities on ZF generation, in this section, we firstly explore the effects of impurities on polarization shielding and on CTEM instability, respectively, in D-T plasmas. The fueling ratios of D and T are set as 50% and 50% according to [29] with the effective ion mass number $A_{i,eff}=2.5$. The typical parameters for CTEM instability in the presence of impurity are used unless otherwise stated: the safety factor $q$=1.8, $\hat{s}$ =0.8, $\varepsilon_0$ =0.18, $T_z=T_i=T_e$ ( $\tau_z=\tau_i=1$ ), $R/L_{Tz}=R/L_{Ti}=2.2$ , $R/L_{Te}=6.9$ , $R/L_{ne}=6.9$ , $k_{\perp 0}\rho_{i,eff}=0.2$ with $\rho_{i,eff}$ being the effective ion gyroradius, $L_{ez}=1$ , $f_c=0.01$ . The tolerance concentration for fully ionized $Be^{4+}$ in ITER are 2% [41], and $f_c=10^{-4}$ for W according to JET [42]. In particular, one special species of impurity in D-T plasmas, the energetic alpha particles $He^{2+}$ from the D-T reaction, mainly heat electrons. Then, the electrons will transfer their energy to ions through the collisional process between electrons and ions. In the present work, we assume $T_i=T_e$ ( $\tau_i=1$ ), $T_e\leq T_z$ ( $\tau_z\leq 1$ ) for high temperature $He^{2+}$ with $f_c=0.1$ [43].

First, we study the variation of the total polarization shielding $g_i\chi_{i,z}+g_z\chi_{z,z}$ due to the inclusion of impurities in Fig. 1. We can see that the total polarization shielding is increased by impurities as shown in Fig. 1(a), and the bigger value of $g_z$, the bigger level of the maximum total polarization shielding. While, in Fig. 1 (b), the total polarization shielding becomes smaller (larger) in the presence of high temperature $He^{2+}$ for $\tau_z <$ (>) 0.5. The dashed black line corresponding to the case without impurity almost coincides with green solid line, which indicates that high temperature He$^{2+}$ with $\tau_z = 0.5$ does not apparently influence the total polarization shielding. It should be noted that $g_i\chi_{i,z} + g_z\chi_{z,z}$ is reduced to $\tau_i\chi_{i,z}$ for the case without impurity ($f_c = 0.00$), and $g_i\chi_{i,z} + g_z\chi_{z,z} - \tau_i\chi_{i,z} = Zf_c(Z\chi_{z,z}\tau_z - \tau_i\chi_{i,z})$. Both $\chi_{i,z}$ and $\chi_{z,z}$ include the classical and neoclassical components as explained in Refs. [10, 16, 44]. Moreover, $\chi_{i,z}$ and $\chi_{z,z}$ become to be comparable when $q_r\rho_{i,eff}$ trends to unity. Thus, $g_i\chi_{i,z} + g_z\chi_{z,z} - \tau_i\chi_{i,z} = Zf_c\chi_{i(z),z}(Z-1) > 0$ for $\tau_i = \tau_z = 1$ as shown in Fig. 1(a), and $g_i\chi_{i,z} + g_z\chi_{z,z} - \tau_i\chi_{i,z} = Zf_c\chi_{i(z),z}(2\tau_z - 1) \leq (>)0$ for high temperature He$^{2+}$ with $\tau_i = 1$ and $\tau_z \leq (>) 0.5$ as shown in Fig. 1(b).

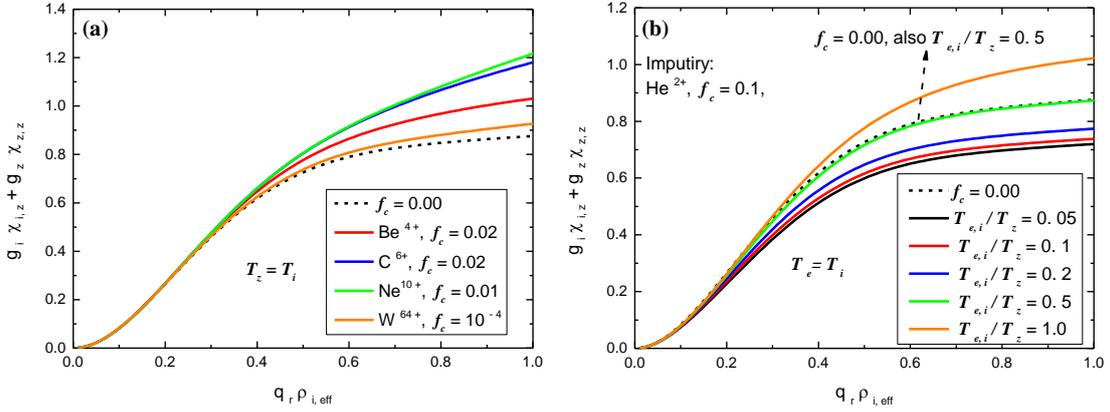

Fig. 1. The total polarization shielding versus $q_r\rho_{i,eff}$. The black dashed lines correspond to the cases without impurity, while the solid lines are cases with impurities.

Now, we come to show the impurity effects on CTEM instability. The impurity density profile is known to have impact on plasma confinement [41]. Thus, we explore how impurity density profile affects the normalized growth rate, frequency and radial group velocity of CTEM instability in Fig. 2. The impurity is chosen as Be$^{4+}$ with concentration around the tolerance value in ITER [41], and $T_z = T_i$ is assumed. $L_{ez} < (>)0$ represents the outwardly (inwardly) peaked impurity density profile. Larger value of $|L_{ez}|$ means steeper impurity profile because $R/L_{ne}$ is fixed. In Fig. 2(a), the normalized growth rate of CTEM is enhanced (reduced) by Be$^{4+}$ with negative (positive) $L_{ez}$, meanwhile, the normalized frequency in Fig.

2(b) is slightly reduced (enhanced) by $Be^{4+}$ with negative (positive) $L_{ez}$. In particular, $\left|\hat{v}_{gr}\right|$ becomes lower by increasing $L_{ez}$ as shown in Fig. 2(c), and $\left|\hat{v}_{gr}\right|$ is enhanced (reduced) by the presence of $Be^{4+}$ with negative (positive) $L_{ez}$. Besides, it also clearly shows that all these impurity effects on CTEM instability are strengthened by the increase of $|L_{ez}|$ and $f_c$, which is consistent with [23]. Finally, we note that impurity influences on $\left|\hat{v}_{gr}\right|$ in Fig. 2(c) is more evident than those on linear growth rate and frequency in Fig. 2(a) and (b).

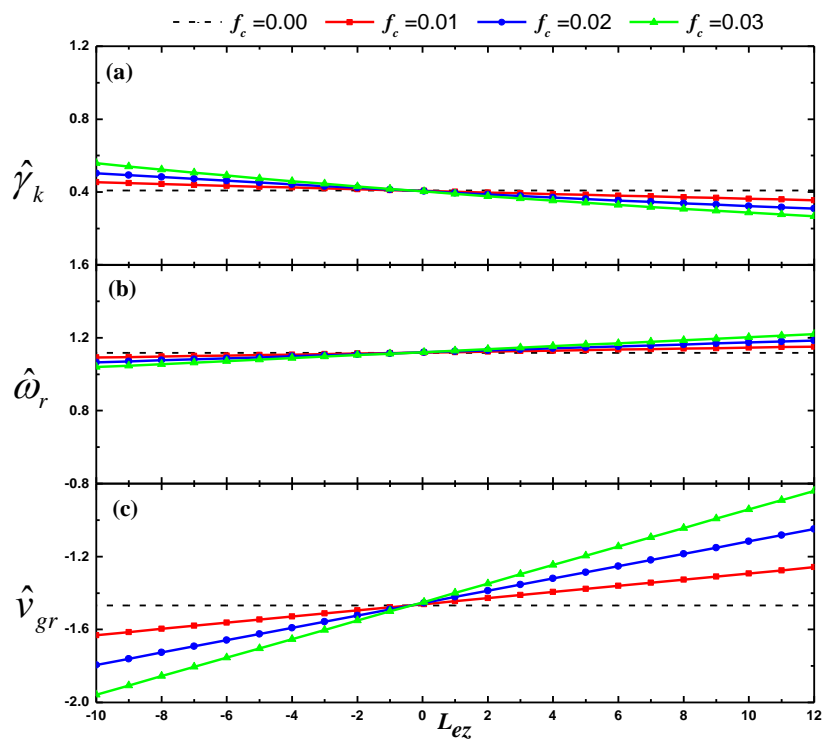

Fig. 2. Normalized growth rate (a), real frequency (b) and radial group velocity (c) of CTEM instability versus $L_{ez}$ with different $f_c$. Fully ionized $Be^{4+}$ is taken as the impurity with $T_z = T_i$. The black dashed, red squared, blue circle and green triangular lines correspond to $f_c=0.00$, $f_c=0.01$, $f_c=0.02$ and $f_c=0.03$, respectively.

In Fig. 3, we explore the impact of electron to impurity temperature ratio on CTEM instability. The impurity is chosen as $Be^{4+}$ with different $f_c$, and $L_{ez}=1$, $T_z = T_i$. In addition, $T_e \geq T_i$ which is suitable for CTEM instability is assumed. In Fig. 3, it is shown that the normalized CTEM growth rate is increased with the increase of $T_e/T_{i,z}$, but the increase of $T_e/T_{i,z}$ weakly reduces the normalized frequency, and the absolute value of the normalized radial group velocity is enhanced by increasing $T_e/T_{i,z}$. Moreover, $\dfrac{\hat{\gamma}_k}{|\hat{v}_{gr}|}$ is finally increased with the increase of $T_e/T_{i,z}$ due to much stronger variation of $\hat{\gamma}_k$, and this will influence the ZF generation. In addition, increasing the concentration of $Be^{4+}$ with $L_{ez}=1$ has very weak influences on the normalized growth rate, frequency and radial group velocity of CTEM instability for fixed temperature ratio.

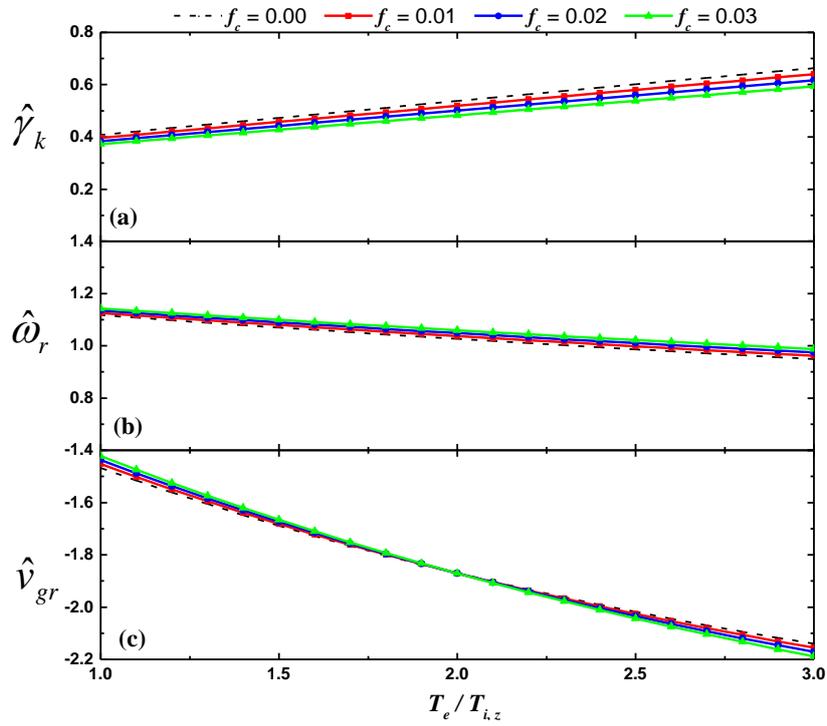

Fig. 3. Normalized growth rate (a), real frequency (b) and radial group velocity (c) of CTEM instability versus electron to impurity/ion temperature ratio with impurity (solid lines) and without impurity (dashed lines). The impurity is fully ionized $Be^{4+}$ with $L_{ez}=1$ and $T_i=T_z$. The black dashed, red squared and blue circle and green triangular lines correspond to $f_c=0.00$, $f_c=0.01$, $f_c=0.02$, $f_c=0.03$ respectively.

Especially, the influences of high temperature $He^{2+}$ from D-T reaction with $f_c=0.1$ on CTEM are shown in Fig. 4. Here, we assume $T_e=T_i\leq T_z$ because of the energetic alpha particles produced by D-T reaction as mentioned before. In the presence of high temperature

He$^{2+}$, both the normalized growth rate in Fig. 4 (a) and $|\hat{v}_{gr}|$ in Fig. 4 (c) are reduced by the increase of $T_{e,i}/T_z$, while the real frequency in Fig. 4 (b) is slightly enhanced by increasing $T_{e,i}/T_z$. Moreover, when $T_{e,i}/T_z < 0.35$, both $\hat{\gamma}_k$ and $|\hat{v}_{gr}|$ are enhanced by the presence of high temperature He$^{2+}$. When $T_{e,i}/T_z$ is further increased to $T_{e,i}/T_z > 0.35$, both $\hat{\gamma}_k$ and $\hat{\omega}_r$ trend to saturate at the level without impurity, but $|\hat{v}_{gr}|$ tends to saturate at a lower level as compared to the case without impurity.

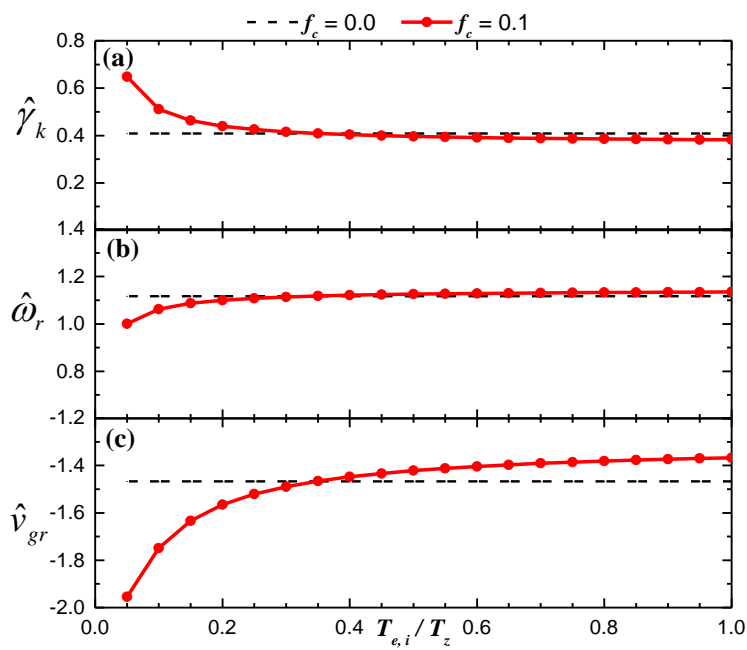

Fig. 4. Normalized growth rate (a), real frequency (b) and radial group velocity (c) of CTEM instability versus electron/ion to impurity temperature ratio (red solid lines) and the case without impurity (black dashed lines). $T_e = T_i$ is assumed. The impurity is high temperature He$^{2+}$ from D-T reaction with $L_{ez} = 1$, $f_c = 0.1$.

## IV. Impact of impurities on ZF growth rate.

In this section, we explore how impurities influence ZF growth rate in D-T plasmas based on the analyzation in section III. Firstly, we investigate the effects of $f_c$ and different kinds of fully ionized non-trace light impurities on ZF growth rate in Fig. 5. The maximum normalized ZF growth rate, $\hat{\gamma}_{ZF,\max}$, is reduced by these impurities. This is mainly because the total polarization shielding (in the denominator of the expression for $\hat{\gamma}_{ZF,\max}$ in Eq. (18)) is enhanced by the presence of impurities, and $\frac{\hat{\gamma}_k}{|\hat{v}_{gr}|}$ (in the numerator of $\hat{\gamma}_{ZF,\max}$ in Eq. (18)) is almost not changed by impurities. The fully ionized impurities with higher $f_c$ and heavier mass result in more serious decrease of $\hat{\gamma}_{ZF,\max}$, which is mainly due to the much stronger enhancement of the total polarization shielding by increasing $g_z$ as shown in Fig. 1 (a). Moreover, the invisible variation of $\frac{\hat{\gamma}_k}{|\hat{v}_{gr}|}$ due to the presence of impurity indicates that the ZF radial wavelength regime corresponding to the growth of ZF is not changed by these impurities.

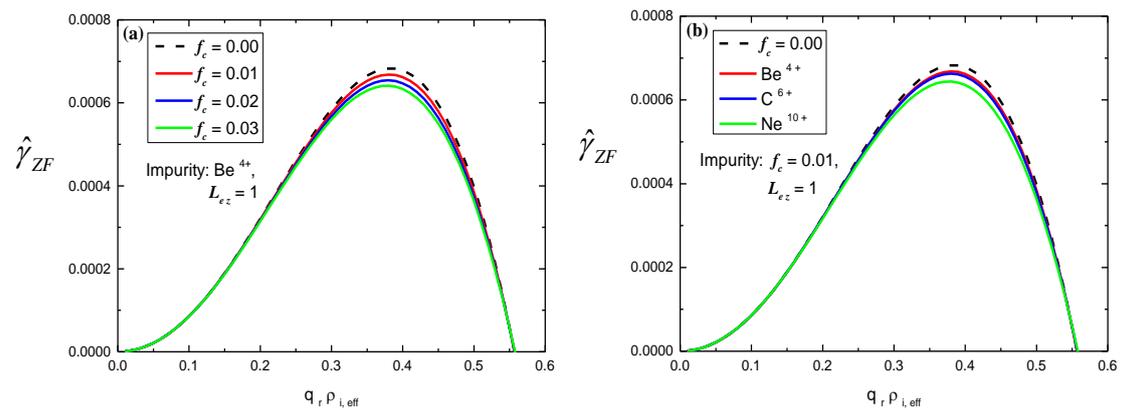

Fig. 5. The normalized ZF growth rate versus $q_r \rho_{i,eff}$ for different concentration of $Be^{4+}$ (a) and different kinds of fully ionized non-trace light impurities (b). The black dashed lines are the case in the absence of impurity and the solid lines correspond to the cases with impurity.

Fig. 6 displays the influences of impurity density profile on the normalized ZF growth rate. As compared to the case without impurity (dashed lines), in Fig. 6 (a), we can see that the effects of impurity density profile on $\hat{\gamma}_{ZF,\max}$ is complicated. It shows that the maximum normalized ZF growth rate can be enhanced by $Be^{4+}$ with relatively steep impurity density profile, i.e., large $|L_{ez}|$ (the sign of $L_{ez}$ can be either positive or negative as stated in section II). This is because impurity effects on CTEM instability become more and more evident with the increase of $|L_{ez}|$ as shown in Fig. 2. The enhancement of $\frac{\hat{\gamma}_k}{|\hat{v}_{gr}|}$ by impurity with relatively steep density profile overcomes that of total polarization shielding, leading to the final enhancement of the maximum normalized ZF growth rate. In addition, the enhancement of $\frac{\hat{\gamma}_k}{|\hat{v}_{gr}|}$ also indicates that the regime of ZF radial scale corresponding to the growth of ZF also becomes wider with the increase of $|L_{ez}|$. While, the presence of $Be^{4+}$ with relatively flat density profile, i.e., small $|L_{ez}|$, reduces the normalized ZF growth rate. For this case, the variation of $\frac{\hat{\gamma}_k}{|\hat{v}_{gr}|}$ is weak, and the enhancement of the polarization shielding by impurity is the main reason causing the reduction of maximum normalized ZF growth rate.

In order to illustrate the complicated results in Fig. 6(a) more clearly, we show the ratio between the maxmium normalized ZF growth rate in the presence of impurity ($\hat{\gamma}_{ZF,\max(z)}$) and that without impurity ($\hat{\gamma}_{ZF,\max(0)}$) as a function of $L_{ez}$ in Fig. 6(b). For $Be^{4+}$, $\frac{\hat{\gamma}_{ZF,\max(z)}}{\hat{\gamma}_{ZF,\max(0)}} > (<) 1$ for relatively steep (flat) impurity density profile, which clearly means that the maximum normalized ZF growth rate is enhanced (reduced) by impurities. But, when changing impurity from $Be^{4+}$ to $C^{6+}$ (their concentrations are the same), the critical value of $|L_{ez}|$ for distinguishing the enhancement and reduction effects of impurity on the maximum normalized ZF growth rate becomes smaller. In addition, increasing the mass of the fully ionized non-trace impurities shows different effects on $\frac{\hat{\gamma}_{ZF,\max(z)}}{\hat{\gamma}_{ZF,\max(0)}}$ in Fig. 6 (b). When $|L_{ez}| < (>) 4$, for a given impurity density profile, changing impurity from $Be^{4+}$ to $C^{6+}$, $\frac{\hat{\gamma}_{ZF,\max(z)}}{\hat{\gamma}_{ZF,\max(0)}}$ is reduced (enhanced). This is mainly because of the much stronger enhancement of total polarization shielding by increasing the impurity weighting factor $g_z$ (from $Be^{4+}$ to $C^{6+}$) when $|L_{ez}|<4$ as shown in Fig. 1 (a). But, for $|L_{ez}|>4$, the enhancement of $\frac{\hat{\gamma}_k}{|\hat{v}_{gr}|}$ is much stronger for $C^{6+}$ than that for $Be^{4+}$, and it overcomes the variation of total polarization shielding from $Be^{4+}$ to $C^{6+}$. Therefore, $\frac{\hat{\gamma}_{ZF,\max(z)}}{\hat{\gamma}_{ZF,\max(0)}}$ is reduced (enhanced) from $Be^{4+}$ to $C^{6+}$ for $|L_{ez}|<(>)4$.

The enhancement of the maximum normalized ZF growth rate by impurity with large positive might bring some good news for the achievement of better confinement by injecting light impurity in the core plasmas. In DIII-D, the flow shearing rate is found to be increased due to the injection of Ne and confinement improvement is hence observed [26]. Our findings on the effects of impurity with relatively steep inwardly peaked density profile could provide a possible explanation for these experimental results. One may argue that the impurity confinement is also improved due to the enhancement of ZF induced by light impurity injection, which is possibly unfavourable for improving confinement. Thus, detailed analysis of the impurity effects on confinement needs further investigations in the future.

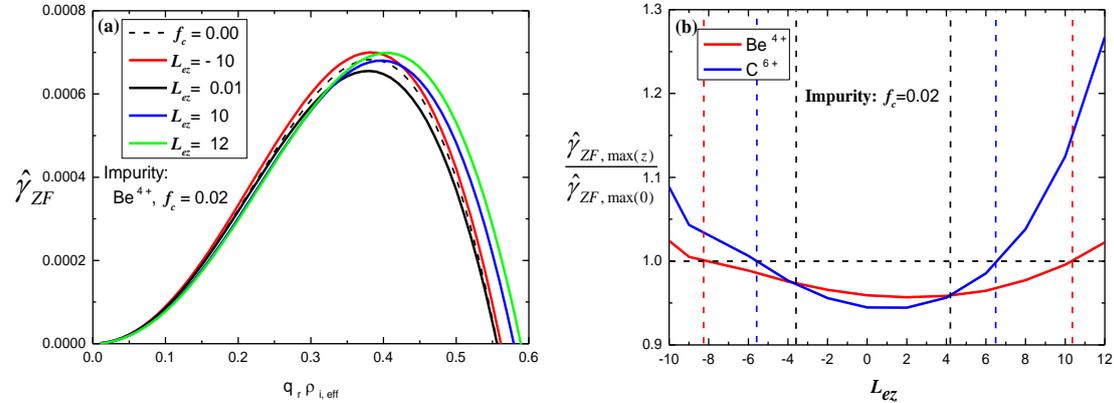

Fig. 6. (a) The normalized ZF growth rate versus $q_r \rho_{i,eff}$ for $Be^{4+}$ (solid lines) with $L_{ez}<(>)0$ corresponding to the outwardly (inwardly) peaked impurity profile and for the case without impurity (black dashed line). (b) The ratio between the maxmium normalized ZF growth rate with and without impurity versus $L_{ez}$.

In Fig. 7, we show that the change of the normalized ZF growth rate versus $q_r \rho_{i,eff}$ with different electron to impurity ($Be^{4+}$) temperature ratio. Compared to the dashed lines corresponding to the cases without impurity, the introduction of Be$^{4+}$ with $T_e/T_{i,z} = 1$ reduces the maximum normalized ZF growth rate. It is mainly due to the enhancement of total polarization shielding by impurity and the invisible variation of $\dfrac{\hat{\gamma}_k}{|\hat{v}_{gr}|}$ when $T_e/T_{i,z} = 1$. While, impurity effects on the maximum normalized ZF growth rate become weak when $T_e/T_{i,z} = 2, 3$. The enhancement of total polarization shielding overcomes the slightly enhancement of $\dfrac{\hat{\gamma}_k}{|\hat{v}_{gr}|}$ leading to the significant reduction of $\hat{\gamma}_{ZF,\max}$ with the increase of $T_e/T_{i,z}$. Thus, the reduction of the maximum normalized ZF growth rate due to the presence of impurity with $T_e/T_{i,z} = 2, 3$ can not be clearly observed. The enhancement of $\dfrac{\hat{\gamma}_k}{|\hat{v}_{gr}|}$ by increasing $T_e/T_{i,z}$ also indicates the wider region corresponding to the growth of ZF.

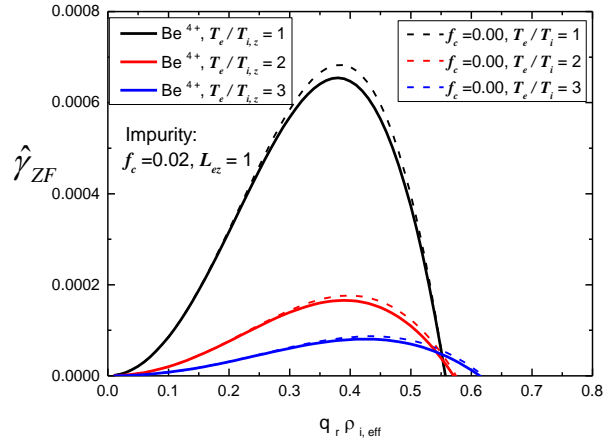

Fig. 7. The normalized ZF growth rate versus $q_r \rho_{i,\,eff}$ for different temperature ratios with (solid lines) and without (dashed lines) impurity.

The obvious reduction of the normalized ZF growth rate by increasing $T_e/T_{i,z}$ reflected by the solid lines in Fig. 7 may ease the suppression of ZF on ambient CTEM, thus, the transport level of impurity driven by CTEM turbulence might become higher. This might be possibly consistent with the results in gyrokinetic simulation [12], experiments [45] and the references therein, where the electron dominated auxiliary heating is widely recognized to be beneficial for transporting impurity.

In particular, we also explore the effects of the temperature ratio between electron and high temperature $He^{2+}$ in Fig. 8. Here, $T_e = T_i$ is assumed. The maximum normalized ZF growth rate is enhanced (reduced) by high temperature $He^{2+}$ with temperature ratios $T_{e,i}/T_z \leq 0.5$ ($T_{e,i}/T_z = 1$). The enhancement of $\frac{\hat{\gamma}_k}{|\hat{v}_{gr}|}$ and the reduction of total polarization shielding by high temperature $He^{2+}$ (as shown in Fig. 1 (b)) together significantly enhance the maximum normalized ZF growth rate when $T_{e,i}/T_z \leq 0.5$. While, for much higher value of $T_{e,i}/T_z$ such as $T_{e,i}/T_z = 1$, the variation of $\frac{\hat{\gamma}_k}{|\hat{v}_{gr}|}$ becomes weak, thus, the reduction of the maximum normalized ZF growth rate by high temperature $He^{2+}$ is mainly due to the enhancement of the total polarization shielding $g_i \chi_{i,z} + g_z \chi_{z,z}$. In order to straightforwardly exhibit the temperature ratio dependence of ZF growth rate, we plot the variation of $\frac{\hat{\gamma}_{ZF,\max(z)}}{\hat{\gamma}_{ZF,\max(0)}}$ as a function of $T_{e,i}/T_z$ in Fig. 8 (b), which monotonically decreases with the increase of $T_{e,i}/T_z$. For $T_{e,i}/T_z < (>) 0.65$, the maximum normalized ZF growth rate is enhanced (reduced) as compared to the case without high temperature $He^{2+}$ impurity.

These results in Fig. 8 indicate that evaluating the impact of high temperature $He^{2+}$ on ZF generation should be comprehensive. Especially, the significant enhancement of ZF growth by higher temperature $He^{2+}$ might have a significant ZF regulation on DW turbulence, which is favorable for confinement improvement. While, the growth rate of CTEM instability is increased by higher temperature $He^{2+}$ as shown in Fig. 4(a). Thus, the accurate predictions of the effects of alpha particles from D-T reaction on ZF and confinement in future burning plasmas during the complicated thermalization and slow down processes need careful investigation.

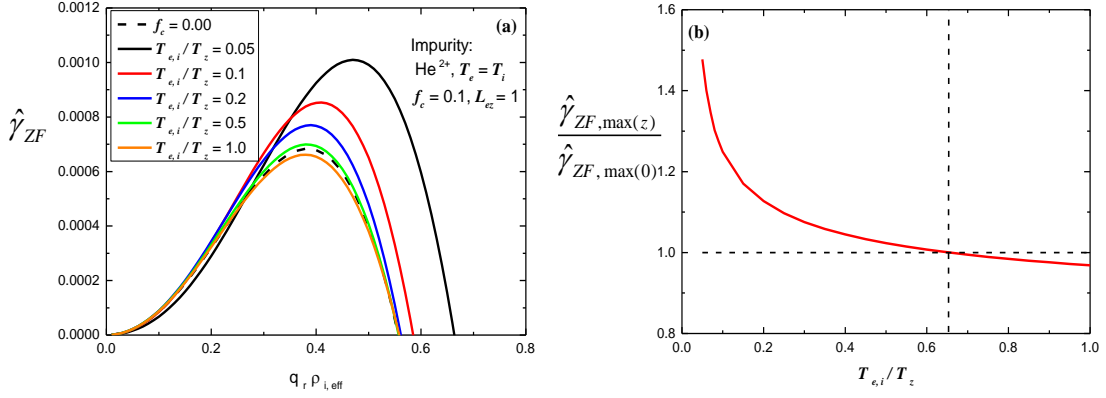

Fig. 8. (a) The normalized ZF growth rate versus $q_r \rho_{i,\,eff}$ for high temperature $He^{2+}$ with different electron to impurity temperature ratios (solid lines) and the case without impurity (black dashed line). (b) The ratio between the maximum normalized ZF growth rate with and without high temperature $He^{2+}$ from D-T reaction versus $T_{e,i}/T_z$. $T_e = T_i \leq T_z$ is assumed.

Finally, we study the effects of different ionized trace W on ZF growth rate. The value of $f_c$ is the order of $10^{-4}$ according to the tolerance concentration in JET [42]. The results are shown in Fig. 9. The presence of W also slightly reduces $\hat{\gamma}_{ZF,\,\max}$ as compared to the case in the absence of W. This result can be easily understood base on the previous analyses. The total polarization shielding can be slightly enhanced by trace W, but the trace W almost does not affect the CTEM instability.

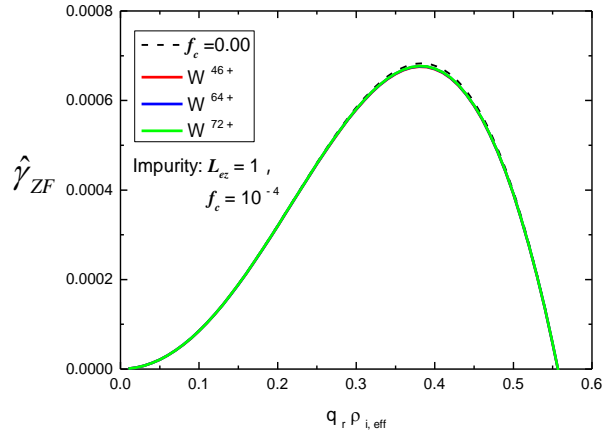

Fig. 9. The normalized ZF growth rate versus $q_r \rho_{i,\,eff}$ for W with different ionized stage (solid line) and the case without impurity (black dashed lines).

## V. Conclusions and discussions.

In the present work, we study the impact of impurities on ZF generation driven by CTEM turbulence in D-T plasmas. Impurity effects on both polarization shielding and CTEM instability are considered. The ZF modulated radial current driven by turbulence is directly calculated, which is verified to be the same as the expression in Refs. [10, 11] in the absence of impurity. The growth rate and real frequency of CTEM instability in the presence of impurity are derived through the bounce kinetic equation for electrons and fluid approximations for ions and impurities. Then, we obtain the analytic expression for ZF growth rate with impurity by balancing the polarization current and the modulated turbulent driven current. This work extends our previous study of the impurity effects on the polarization shielding of ZF, i.e., residual (or RH) ZF [16]. The principal results of the present work are summarized in table 1.

| Types of impurities | Impurity effects on $g_i \chi_{i,z} + g_z \chi_{z,z}$ | Impurity effects on $\dfrac{\hat{\gamma}_k}{|\hat{v}_{gr}|}$ | Impurity effects on $\hat{\gamma}_{ZF,\max}$ |
|---|---|---|---|
| Fully ionized non-trace light impurities | Increase | Small $|L_{ez}|$: weak | Small $|L_{ez}|$: decrease |
|  |  | Large $|L_{ez}|$: increase | Large $|L_{ez}|$: increase |
| $He^{2+}$ from D-T reaction with $T_e/T_z \leq 1$ | $T_e/T_z < 0.5$: decrease | $T_e/T_z < 0.2$: increase | $T_e/T_z < 0.65$: increase |
|  | $T_e/T_z > 0.5$: increase | $T_e/T_z \geq 0.2$: weak | $T_e/T_z > 0.65$: decrease |
| Trace W | Slightly increase | Weak | Slightly decrease |

Table 1. Overview of the main results in this work.

As we know, various impurities could be present in ITER. Our results listed in Table 1 showed that these impurities, including the fully ionized non-trace light impurities, $He^{2+}$ from D-T reaction, and highly ionized W, do affect the generation of ZF driven by CTEM turbulence based on the set of parameters taken in section III. The parametric dependence of ZF generation in CTEM turbulence has been found in gyrokinetic simulations [7, 46, 47]. Thus, the conclusions in the present work are probably also parametric dependence, which may be worth further investigation. These results might have some possible relevance to evaluate impurity effects on plasma confinement and transport. For example, the presence of non-trace light impurities with steep inwardly peaked density profile can enhance the normalized ZF growth rate, which may indicate the improvement of confinement, and might be qualitatively consistent with the experimental results in DIII-D [21]. The possible significance of the effects of high temperature $He^{2+}$ from D-T reaction on ZF generation is also discussed in detail in the context.

Finally, we note that the monochromatic wave packet was used in the present work. When disparate scale interactions are dominant, a turbulence fluctuation spectrum of the form $\left[n(k)\right]^2 \sim k^{-3}/\left(1+k^2\right)^3$ was reported in [48], which is in fair agreement with experimental results in Tore Supra [34]. Thus, it is very interesting to investigate the impurity effects on ZF generation using the non-monochromatic spectrum. In addition, the assumption of spatiotemporal scale separation is usually required by wave kinetic equation approach [11, 49] and renormalized statistical theory [32] to enable the analytical derivations. However, the ZF radial scale corresponding to the maximum ZF growth rate in this work is a little bit smaller than the DW radial scale. The similar limitation also existed and was discussed in [10, 32, 49]. Actually, fine scale ZF (comparable to or smaller than DW scale) has been observed in many gyrokinetic simulation works [7, 8, 50] as well as in JET experiments [51]. More important role of temporal scale separation than that of spatial scale separation is pointed out in [52] via comparing the two methods of the wave kinetic equation and the coherent mode coupling [53]. In the present work, the validity of temporal scale separation is well satisfied by estimating $\frac{\gamma_{ZF}}{\gamma_k} \sim \frac{\hat{\gamma}_{ZF}}{\hat{\gamma}_k} \frac{L_{ne}}{c_s} \frac{v_{thi}}{R} \frac{R^2}{\rho_i^2} \frac{\rho_i^2}{L_{ne}^2} \sim \frac{\hat{\gamma}_{ZF}}{\hat{\gamma}_k} \frac{R}{L_{ne}} \sim 10^{-3}$ to $10^{-2}$. On the other hand, modulational instability based on wave kinetic equation approach may not be perfect for fine scale ZF generation, it may be also worth investigating other mechanisms to generate fine scale ZF such as coherent mode coupling method [53]. Moreover, there are few works considering EM effects on the DW-ZF dynamical system with impurity expect the study of EM effects on impurity transport [54]. Extending our present work to ZF and zonal fields [1, 55, 56] in EM turbulence in the future is also worthwhile. The last but not the least, this work only focused on how impurity affects ZF linear growth rate. Impurity effects on the nonlinear evolution of ZF are still open question and should also be investigated in the future.

## Acknowledgments

We are grateful to P. H. Diamond for his heuristic discussions. We also thank J.Q. Dong and X. Q. Xu for helpful suggestions and discussions. The work was supported by the NSFC Grant Nos. 11675059 and 11305071, the MOST of China under Contract No. 2013GB112002.